\documentclass{article}

\usepackage{arxiv}
\usepackage{comment}
\usepackage[utf8]{inputenc} 
\usepackage[T1]{fontenc}    
\usepackage{hyperref}       
\usepackage{url}            
\usepackage{booktabs}       
\usepackage{amsfonts}       
\usepackage{nicefrac}       
\usepackage{microtype}      
\usepackage{lipsum}		
\usepackage{graphicx}
\usepackage[square, numbers]{natbib}
\usepackage{doi}
\usepackage{pifont}
\usepackage{makecell}
\usepackage{tabularx}
\usepackage{lscape}
\usepackage{subfig}
\usepackage{adjustbox}

\usepackage{graphicx,color}
\usepackage{hyperref}
\usepackage{pdfpages} 
\usepackage{media9}
\usepackage{comment}
\usepackage{latexsym}
\usepackage{bm}
\usepackage{xspace}
\usepackage{booktabs}
\usepackage{threeparttable}

\title{A Summary of the ComParE COVID-19 Challenges}

\author{Harry Coppock\thanks{\texttt{harry.coppock@imperial.ac.uk}}\\
	Imperial College London, UK\\
	\And
     Alican Akman \\
	Imperial College London, UK\\
    \And 
    Christian Bergler\\
    FAU Erlangen-Nürnberg, Germany\\
    \And
    Maurice Gerczuk \\
    Universität Augsburg, Germany\\
    \And
    Chloë Brown\\
    University of Cambridge, UK\\
    \And
    Jagmohan Chauhan\\
    University of Southampton, UK\\
    \And
    Andreas Grammenos\\
    University of Cambridge, UK\\
    \And
    Apinan Hasthanasombat\\
    University of Cambridge, UK\\
    \And
    Dimitris Spathis\\
    University of Cambridge, UK\\
    \And
    Tong Xia\\
    University of Cambridge, UK \\
    \And
    Pietro Cicuta\\
    University of Cambridge, UK\\
    \And
    Jing Han\\
    University of Cambridge, UK\\
    \And
    Shahin Amiriparian\\
    Universität Augsburg, Germany\\
    \And
    Alice Baird\\
    Universität Augsburg, Germany\\
    \And
    Lukas Stappen\\
    Universität Augsburg, Germany\\
    \And
    Sandra Ottl\\
    Universität Augsburg, Germany\\
    \And
    Panagiotis Tzirakis\\
    Imperial College London, UK\\
    \And
    Anton Batliner\\
    Universität Augsburg, Germany\\
    \And
    Cecilia Mascolo\\
    University of Cambridge, UK\\
    \And
    Björn W. Schuller \\
	Imperial College London, UK \\ and University of Augsburg, Germany\\

}

\renewcommand{\shorttitle}{ComParE COVID-19 Challenges}

\hypersetup{
pdftitle={A template for the arxiv style},
pdfsubject={q-bio.NC, q-bio.QM},
pdfauthor={David S.~Hippocampus, Elias D.~Striatum},
pdfkeywords={First keyword, Second keyword, More},
}

\begin{document}
\maketitle

\begin{abstract}
	The COVID-19 pandemic has caused massive humanitarian and economic damage. Teams of scientists from a broad range of disciplines have searched for methods to help governments and communities combat the disease. One avenue from the machine learning field which has been explored is the prospect of a digital mass test which can detect COVID-19 from infected individuals' respiratory sounds. We present a summary of the results from the INTERSPEECH 2021 Computational Paralinguistics Challenges: COVID-19 Cough, (CCS) and COVID-19 Speech, (CSS). 
\end{abstract}

\keywords{COVID-19 \and Machine Learning \and Mass Test \and Challenge \and Speech \and Cough}

\section{Introduction}
Significant work has been conducted exploring the possibility that COVID-19 yields unique audio biomarkers in infected individuals' respiratory signals \cite{brown_exploring_2020,xia_uncertainty-aware_2021, imran_ai4covid-19_2020, Sharma2020, bagad2020cough, pinkas_sars-cov-2_2020, orlandic_coughvid_2020, Perez_deep_cough, pizzo2021iatos, QianMSC-COVID, bartl-pokorny_voice_2020, CoppockCOVID2021, minaDetecting, ponomarchuk2022}. This has shown promising results although many still remain sceptical, suggesting that models could simply be relying on spurious bias signals in the datasets \cite{coppock2021SevenGrainsofSalt,CoppockCOVID2021}. These worries have been supported by findings that when sources of bias are controlled, the performance of the classifiers decreases \cite{han2021sounds, coppock2021SevenGrainsofSaltcorrespondence}. Along with this, cross dataset experiments have reported a marked drop in performance when models trained on one dataset are then evaluated on another dataset, suggesting dataset specific bias \cite{akman2021eval}.

Last summer, the machine learning community were called upon to address some of these challenges, and help answer the question whether a digital mass test was possible, through the creation of two COVID-19 challenges within the Interspeech Computational Paralinguistics challengE (ComParE) series: COVID-19 Cough, (CCS) and COVID-19 Speech, (CSS)\cite{schuller2021interspeech}. Contestants were tasked to create the best performing COVID-19 classifier from user cough and speech recordings. We note that another COVID-19 detection from audio challenge was run at a similar time to ComParE, named DiCOVA \cite{muguli2021dicova}, and point the inquisitive reader to their blog post \footnote{\href{https://dicova2021.github.io}{https://dicova2021.github.io}} which details a summary of the results.

\section{Challenge Methodology}

Both COVID-19 cough and speech challenges were binary classification tasks. Given an audio signal of a user coughing or speaking, challenge participants were tasked with predicting whether the respiratory signal came from a COVID-19 positive or negative user.  After signing up to the challenge, teams were sent the audio files along with the corresponding labels for both the training and development set. Teams were also sent the audio files from the test set without the corresponding labels. Teams were allowed to submit five predictions for the test set from which the best score was taken. The number of submissions was limited to avoid overfitting to the test set.

\begin{figure}
	\centering
	\subfloat[]{\label{timesub}\includegraphics[width=0.45\textwidth]{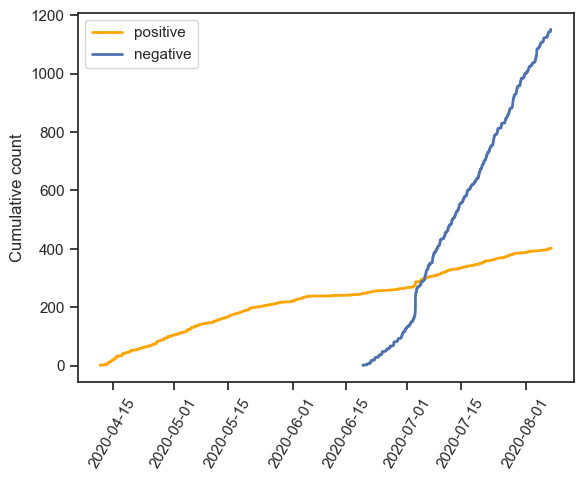}} 
    \subfloat[]{\label{age}\includegraphics[width=0.5\textwidth]{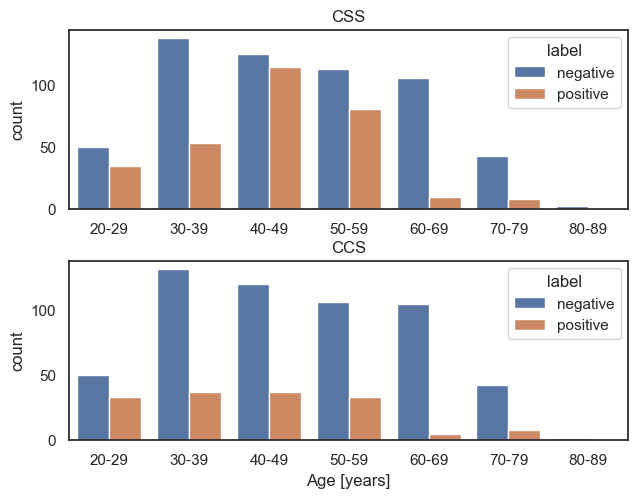}}
    \caption{Figure \ref{timesub} is a cumulative plot detailing when COVID-19  positive and negative submission to both the CCS and CSS were made. Figure \ref{age} details the age distribution of COVID-19  positive and negative participants for the CCS and CSS Sub-Challenges}

	\label{fig:meta}
\end{figure}

The datasets used in these challenges are two curated subsets of the crowd sourced Cambridge COVID-19 Sounds database \cite{brown_exploring_2020,tong2021datasets}. COVID-19 status was self reported and determined through either a PCR or rapid antigen test. The number of samples of both positive and negative cases for these selected subsets are detailed in Table \ref{tab:dataset}. The submission date for both COVID-19 positive and negative case recordings are detailed in Figure \ref{timesub}. Figure \ref{age} shows the age distribution for both CSS and CCS challenges.

\begin{table}[th]
  \caption{ComParE COVID-19 Sub-Challenges dataset splits. Values specify the number of audio recordings. We note that disjoint participant train, development, and test splits were ensured.}
  \label{tab:compare}
  \centering
  \begin{tabular}{ c | c | c | c | c | c | c}
    \toprule
     & \multicolumn{3}{c}{CCS*} &  \multicolumn{3}{c}{CSS $\dagger$}\\
    \cline{2-7}
    \# & Train & Dev & Test & Train & Dev & Test \\
    \hline
    COVID-19-positive & 71 & 48& 39 & 72& 142 & 94 \\
    COVID-19-negative& 215& 183& 169 & 243& 153 & 189 \\
    \hline
    Total& 286 & 231 & 208& 315& 295& 283 \\

    \bottomrule
  \end{tabular}
\label{tab:dataset}

\footnotesize{*CCS -- COVID-19 Cough Sub-Challenge}\\
\footnotesize{$\dagger$ CSS -- COVID-19 Speech Sub-Challenge}
\end{table}

\section{Overview of Methodologies Used in Accepted Papers at Interspeech 2021}
Last year, 44 teams registered in both the ComParE COVID-19 Cough Sub-Challenge (CCS) and the COVID-19 Speech Sub-Challenge (CSS) of which 19 submitted test set predictions. Five of the 19 teams submitted papers to INTERSPEECH which were then accepted. Results for both CCS and CSS were reported in two of these papers, while two papers reported results exclusively for CCS and one paper exclusively for CSS. In this section, we provide a brief overview of methodologies used in these accepted works which included data augmentation techniques, feature types, classifier types, and ensemble model strategies. Teams that did not have their work accepted at INTERSPEECH 2021 will be named NN\_X to preserve anonymity. NN refers to \textit{nomen nescio} and X is the order in which they appear in Figure \ref{fig:resultscough}.The performance measured in Unweighted Average Recall (UAR) achieved by these methodologies is summarised in Table \ref{tab:teams_overview}; UAR has been used as a standard measure in the Computational Paralinguistics Challenges at Interspeech since 2009 \cite{Schuller09-TI2}. It is the mean of the diagonal of the confusion matrices in percent and by that, fair towards sparse classes.Note that UAR is sometimes called `macro-average', see \cite{Manning09-AIT}.

\begin{table}
	\caption{Summary of methodologies used in accepted papers at Interspeech 2021}
	\centering
	\begin{tabular}{lllllll}
		\toprule
		Team Name  & \makecell{Data \\ Aug.} & Feature Type & Classifiers & Ensemble &  \makecell{Cough \\ UAR [\%]} &  \makecell{Speech\\ UAR [\%]}\\
		\midrule
		\textit{Solera-Urena, et al.}  \cite{soleraurena21_interspeech}  & \ding{55}  & \makecell{TDNN-F, \\ VGGish, PASE+} & SVM & \ding{51} & 69.3 & --   \\
		\textit{Casanova, et al.}  \cite{casanova21_interspeech}   & \ding{51} & \makecell{MFCC, \\ mel-spectrogram}  & \makecell{SpiraNet, CNN14, \\ ResNet-38, MobileNet} & \ding{51} & \textbf{75.9} & 70.3   \\
	\textit{Klumpp, et al.} \cite{klumpp21_interspeech}    & \ding{51} & mel-spectrogram  & \makecell{CNN, LSTM, \\ SVM, LR} & \ding{55} & -- & 64.2   \\
		\textit{Illium, et al.} \cite{illium21_interspeech}    & \ding{51} & mel-spectrogram  & Vision Transformer & \ding{55} & 72.0 & --   \\
		\textit{Baseline} \cite{schuller2021interspeech}& \ding{55} & \makecell{openSMILE, \\ openXBOX, DiFE, \\ DeepSpectrum,\\ auDeep}  & SVM, End2You & \ding{51} & 73.9 & \textbf{72.1}   \\
		\bottomrule
	\end{tabular}
	\label{tab:teams_overview}
\end{table}

\subsection{Data Augmentation}

To combat the limited size and imbalance of the Cambridge COVID-19 Sounds database, the majority of the teams used data augmentation techniques in their implementation. Team \textit{Casanova, et al.} exploited a noise addition method and SpecAugment to augment the challenge dataset~\cite{casanova21_interspeech}. Team \textit{Illium, et al.} targeted spectrogram-level augmentations with temporal shifting, noise addition, SpecAugment and loudness adjustment~\cite{illium21_interspeech}. Instead of using a data augmentation method to manipulate the challenge dataset, team \textit{Klumpp, et al.} used three auxiliary datasets in different languages aiming their deep acoustic model to better learn the properties of healthy speech \cite{klumpp21_interspeech}.

\subsection{Feature Type}

The teams chiefly used spectrogram-level features including mel-frequency cepstral coefficients (MFCC) and mel-spectrograms. For higher-level features, the teams used the common feature extraction toolkits openSMILE \cite{eybenopensmile}, openXBOX \cite{schmittxbow}, DeepSpectrum \cite{shahindeepspec}, and auDeep \cite{freitagauDeep}, where a simple support vector machine (SVM) model was built on top of these features. Team \textit{Solera-Urena, et al.} exploited transfer learning to extract feature embeddings by using pre-trained TDNN-F\cite{VILLALBA2020101026}, VGGish\cite{hershey2017cnn}, and PASE+\cite{ravanelli2020multitask} models with appropriate fine-tuning on the challenge dataset. Team \textit{Klumpp, et al.} targeted to extract their own phonetic features by using an acoustic model consisting of convolutional neural network (CNN) and long short-term memory (LSTM) parts.

\subsection{Classifier Type}

Team \textit{Solera-Urena, et al.} \cite{soleraurena21_interspeech} and the challenge baseline \cite{schuller2021interspeech} fitted a SVM model to high level audio embeddings extracted using TDNN-F\cite{VILLALBA2020101026}, VGGish\cite{hershey2017cnn}, and PASE+\cite{ravanelli2020multitask} models, and the openSMILE framework \cite{eybenopensmile}, respectively. While the challenge baseline \cite{schuller2021interspeech} searched for the complexity parameter of the SVM ranging from $10^{-5}$ to $1$, team \textit{Solera-Urena, et al.} \cite{soleraurena21_interspeech} explored different kernels (linear, RBF), data normalisations (zero mean and unit variance, [0,1] range) and class balancing methods (majority class downsampling, class weighting). In addition to the SVM model, the baseline explored using the multimodel profiling toolkit End2You\cite{tzirakis2018end2you} to train a recurrent neural network using Gated Recurrent Units (GRUs) with hidden units of $64$. Team \textit{Casanova, et al.}  \cite{casanova21_interspeech} utilised the deep models: SpiraNet~\cite{casanova-etal-2021-deep}, CNN14~\cite{kong2020panns}, ResNet-38~\cite{kong2020panns}, and MobileNetv1~\cite{kong2020panns} where they explored kernel size, convolutional dilatation, dropout, number of fully connected layer neurons, learning rate, weight decay and optimizer. Team \textit{Klumpp, et al.} \cite{klumpp21_interspeech} trained SVM and logistic regression (LR) models to perform COVID-19 classification on top of phonetic features extracted by their deep acoustic model. They explored the complexity parameter of the SVM ranging from $10^{-4}$ to $1$. Team \textit{Illium, et al.}  \cite{illium21_interspeech} adapted a vision transformer \cite{dosovitskiy2021image} for mel-spectrogram representations of audio signals. Tree-structured Parzen Estimator-algorithm (TPE) \cite{DBLP:journals/corr/abs-1907-10902} was exploited in \cite{illium21_interspeech} for hyperparameter search mainly exploring embedding size, learning rate, batch size, dropout, number of heads and head dimension. The teams \textit{Solera-Urena, et al.}, \textit{Casanova, et al.}, and the baseline also reported classification results by using the fusion of their best features and classifiers.

\begin{figure}[h]
	\centering
    \includegraphics[width=0.9\textwidth]{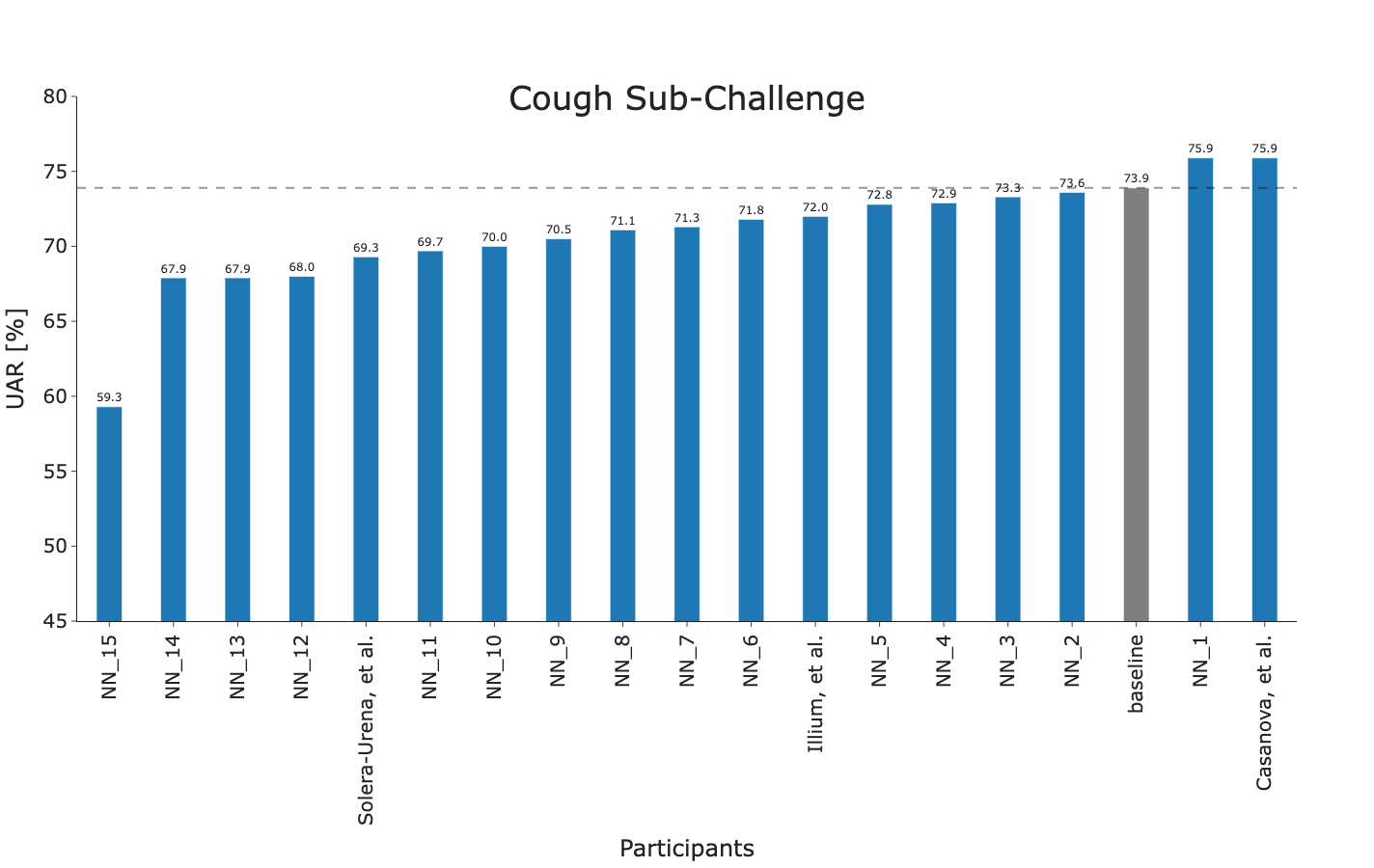}
	\caption{Team performance on the held out test set for the COVID-19 Cough Sub-Challenge. }
	\label{fig:resultscough}
\end{figure}

\begin{figure}[h]
	\centering
    \includegraphics[width=0.9\textwidth]{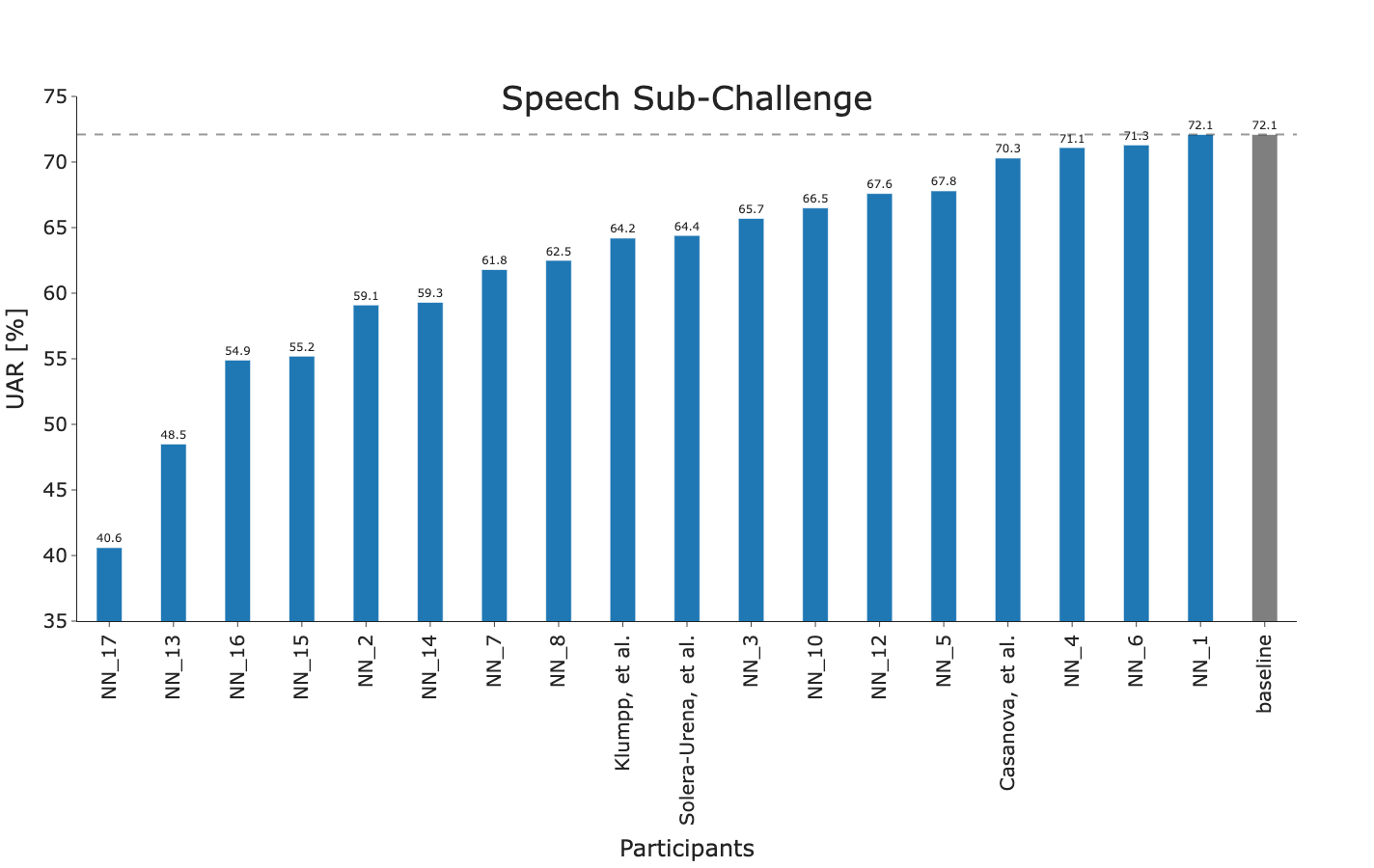}
	\caption{Team performance on the held out test set for the COVID-19 Speech Sub-Challenge.}
	\label{fig:resultsspeech}
\end{figure}

\subsection{Assessment of Performance Measures}

\begin{figure}[!t]
    \centering
    \subfloat[]{\label{coughstat}\includegraphics[width=0.385\textwidth]{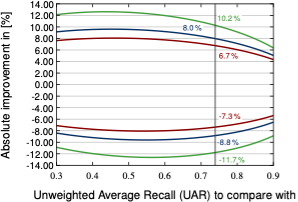}} \hspace{0.1cm}
    \subfloat[]{\label{speechstat}\includegraphics[width=0.5\textwidth]{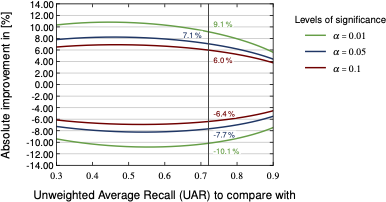}}
    \caption{Two-sided significance test on the COVID-19 Cough (\ref{coughstat}) and Speech (\ref{speechstat}) Test sets with various levels of significance according to a two-sided Z-test.} 
    \label{fig:signif}
\end{figure}

Figure~\ref{fig:signif} visualises a two-sided significance test (based on a Z-test concerning two proportions, \cite{Isaac:2015}, section 5B) employing  the CCS and CSS Test sets and the corresponding baseline systems \cite{schuller2021interspeech}. 
Various levels of significance ($\alpha$-values) were used for   calculating  an absolute deviation with respect to the Test set, being considered as significantly better or worse than the  baseline systems.
Due to the fact that a two-sided  test is employed, the $\alpha$-values must be halved to derive the respective $Z$-score used to calculate the $p$-value of a model fulfilling statistical significance for both sides \cite{Isaac:2015}. Consequently, significantly outperforming the best CCS baseline system (73.9\,\% and 208 Test set samples) at a significance level of $\alpha\,=\,0.01$ requires at least an absolute improvement of $6.7\,\%$; for CSS  (best baseline system with 72.1\,\% and 283 Test set samples), the improvement  required is  $6.0\,\%$. Note that Null-Hypothesis-Testing with $p$-values as criterion has been criticised from its beginning; see the statement of the American Statistical Association in \cite{Wasserstein16-TAS}  and \cite{Batliner20-EAG}.Therefore, we provide this plot with $p$-values as a service for readers interested in this approach, not as a guideline for deciding between approaches.

Another way of assessing performance measures as for their `uncertainty' is computing confidence intervals (CIs). 
 \cite{schuller2021interspeech} employed two different CIs: 
first,  1000x bootstrapping for Test (random selection with replacement) and  UARs based on the same model that was trained with Train and Dev; in the following, the CIs for these UARs are given first. Then, 100x bootstrapping for the corresponding combination of Train and Dev; the different models obtained from these combinations were employed to get UARs for Test and subsequently, CIs; these results are given in second place. Note that for this type of CI, the Test results are often above the CI, sometimes within and in a few cases below, as can be seen in \cite{schuller2021interspeech};  obviously, reducing the variability of the samples in the training phase with bootstrapping results on average in somehow lower performance. For CCS with a UAR of 73.9\,\%, the first CI was   66.0\,\%-82.6\,\%; the second one could not be computed because this UAR is based on a fusion of different classifiers. For CSS with a UAR of 72.1\,\%, the CIs were  66.0\,\%-77.8\,\%  and  70.2\,\%-71.1\,\%, respectively. Both Figure~\ref{fig:signif} and the spread of the CIs reported demonstrate the uncertainty of the results, caused by the relatively low number of data points in the test set.

\section{Results and Discussion}
Figures \ref{fig:resultscough} and \ref{fig:resultsspeech} detail the rankings for the 19 teams which submitted predictions for the test set. We congratulate \cite{casanova21_interspeech} for winning the COVID-19 Cough Sub-Challenge with an UAR of 75.9\,\% on the held out test set. We note that for the COVID-19 Speech Sub-Challenge, no team  exceeded the performance of the baseline which scored 72.1\,\% UAR on the held out test set. To significantly outperform the baseline system for the cough modality, with a significance level of $\alpha = 0.1$, as detailed in Figure \ref{fig:signif}, would require an improvement of 6.7\,\%, an improvement which the winning submission fell short of by 4.7\,\%.

For both Sub-Challenges, teams struggled to outperform the baseline. Postulating why this could be the case one could suggest one, or a combination, of the following: COVID-19 detection from audio is a particularly hard task, the baseline score -- being already a fusion of several state-of-the-art systems for CCS -- represents a performance ceiling and that higher classification scores are not possible for this dataset, or, as a result of the limited size of the dataset, the task lends itself to less data hungry algorithms, such as the openSMILE-SVM baseline models for CSS.

It is important to analyse the level of agreement of COVID-19 detection between participant submissions. This is shown schematically in Figures \ref{fig:agreementcough} and \ref{fig:agreementspeech}. From these figures, we can see that there are clearly COVID-19 positive cases which teams across the board are able to correctly predict, but there are also positive COVID-19 cases which all teams have missed. These findings are reflected in the minimal performance increase of 0.3\,\% and 0.8\,\% for cough and speech tasks, respectively, obtained when fusing $n$ best submission predictions through majority voting schemes. The results from fusing $n$ best models using majority voting are detailed in Figures \ref{fig:fusioncough} and \ref{fig:fusionspeech}. This suggests that models from all teams are depending on similar audio features when predicting COVID-19 positive cases.

Figures \ref{fig:agreementcough} and \ref{fig:agreementspeech} b and c detail the level of agreement across submissions for curated subset of the test set, where participants were selected if they were displaying at least one symptom (b) and when they were displaying no symptoms (c). These figures can be paired with Figure \ref{fig:control} which details the recall scores for positive cases across these same curated test sets. From this analysis, it does not appear that there was a trend across teams to perform favourably on cases where symptoms were being displayed or visa versa. While this does not disprove worries that these algorithms are simply cough or symptom identifiers, it does not add evidence in support of this claim.

\subsection{Limitations}

While this challenge was an important step in exploring the possibilities of a digital mass test for COVID-19, it has a number of limitations. A clear limiting factor of the challenge was the small size of the dataset. While many participants addressed this through data augmentation and regularisation techniques, it restricted the extent to which conclusions could be taken from the results, particularly investigating teams' performance on carefully controlled subsets of the data. We look forward to the newly released COVID-19 sounds dataset \cite{tong2021datasets} which represents a vastly greater source of COVID-19 samples.

A further limitation of this challenge  is the unforeseen correlation between low sample rate recordings, below 12\,kHz, and COVID-19 status. In fact all low sample rate recordings in the challenge for both CCS and CSS were COVID-19 positive. For CCS and CSS there were 30 and 37 low sample rate cases,  respectively. The reason for this is that at the start of the study the label in the survey for COVID-19 negative was unclear, and could have been interpreted as either `not tested' or `tested negative'. For this reason no negative samples from the time period were used. This can be seen in \ref{timesub}. This early version of data collection also correlated with the study allowing for lower sample rate recordings, a feature which later was changed to restrict submissions to higher sample rates. This resulted in all the low sample rate recordings being COVID-19 positive. As can be seen in Figures \ref{fig:agreementcoughsampleratecough}, \ref{fig:agreementspeechsampleratecough}, \ref{fig:controlsampleratecough} and \ref{fig:controlsampleratespeech}, teams' trained models were able to pick up on the sample rate bias, with most teams correctly predicting all the low sample rate cases as COVID-19 positive. When this is controlled for and low sample rate recordings are removed from the test set, as shown in Figures \ref{fig:controlsampleratecough} and \ref{fig:controlsampleratespeech}, teams' performances drop significantly. For the challenge baselines this too was the case, with the fusion of baseline models for CCS falling from 73.8\% to 68.6\% UAR and the opensmile-SVM baseline for CSS dropping from 72.1\% to 70.9\% UAR. This is a great example of the effect of overlooked bias which expresses itself as an identifiable audio feature, leading to inflated classification scores.  We regret that this was not found earlier.

As with most machine learning methods, it still remains unclear how to interpret the decision making process at inference time. This results in it being tricky to determine which acoustic features the model is correlating with COVID-19. Whether that be true, acoustic features caused by the COVID-19 infection or other acoustic bias is still an unanswered question \cite{coppock2021SevenGrainsofSalt}. We also note that this is a binary classification task, in that models only had to decide between COVID-19 positive or negative. This `closed word fallacy' \cite{Batliner20-EAG} leads to inflated performance as models are not tasked with discerning between confounding symptoms such as heavy cold or asthma. Tasking models to predict COVID-19 out of a wide range of possible conditions/symptoms would be a harder task.

In this challenge, participants were provided with the test set recordings (without the corresponding labels). In future challenges, test set instances should be kept private, requiring participants to submit trained models along with pipeline scripts for inference. Teams' test set predictions can then be run automatically by the challenge organisers. This will help in reducing the possibility of overfitting and foul play. We note that there was no evidence of foul play, e.\,g., training in an unsupervised manner on the test set, in this challenge.

Another limitation of this challenge was the lack of meta data that organisers could provide to participants. This tied teams' hands to some extent in evaluating for themselves the level of bias in the dataset and so their opportunity to implement methods to combat it. This was not a desired feature. However, we now point teams towards the newly open sourced COVID-19 Sounds database \cite{tong2021datasets} which also provides collected meta data. It is this dataset from which a subset of samples was taken for this challenge.

\section{Conclusion}

This challenge demonstrated that there is a signal in crowdsourced COVID-19 respiratory sounds that allows for machine learning algorithms to fit a classifier which achieves moderate detection rates of COVID-19 in infected individuals' respiratory sounds. Exactly what this signal is, however, still remains unclear. Whether these signals are truly audio biomarkers in respiratory sounds of infected individuals uniquely caused by COVID-19 or rather identifiable bias in the datasets, such as confounding flu like symptoms, is still an open question to be answered next.

\begin{landscape}
\begin{figure}
	\centering
	\subfloat[]{\label{sublable1}\includegraphics[width=1.4\textheight]{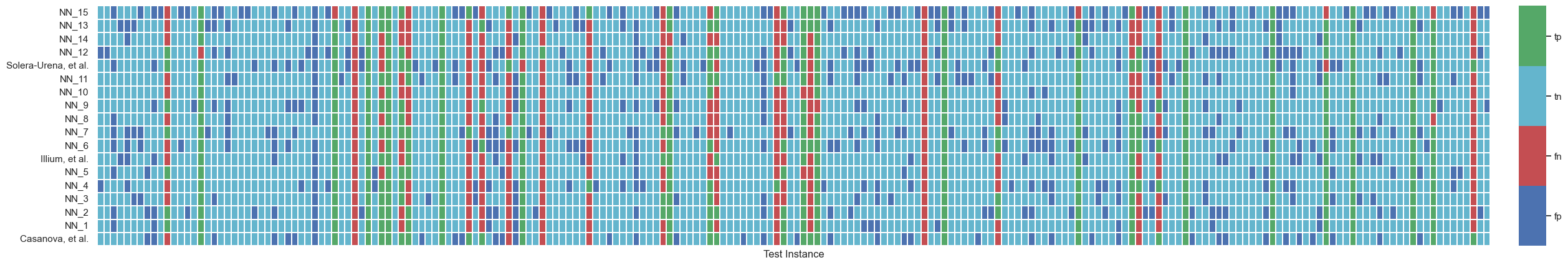}} \\
    \subfloat[]{\label{sublable2}\includegraphics[width=1.4\textheight]{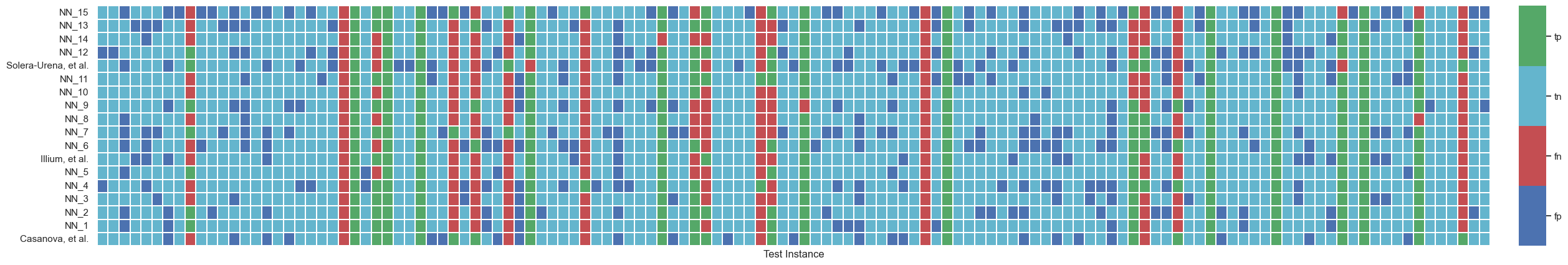}} \\
    \subfloat[]{\label{sublable3}\includegraphics[width=1.4\textheight]{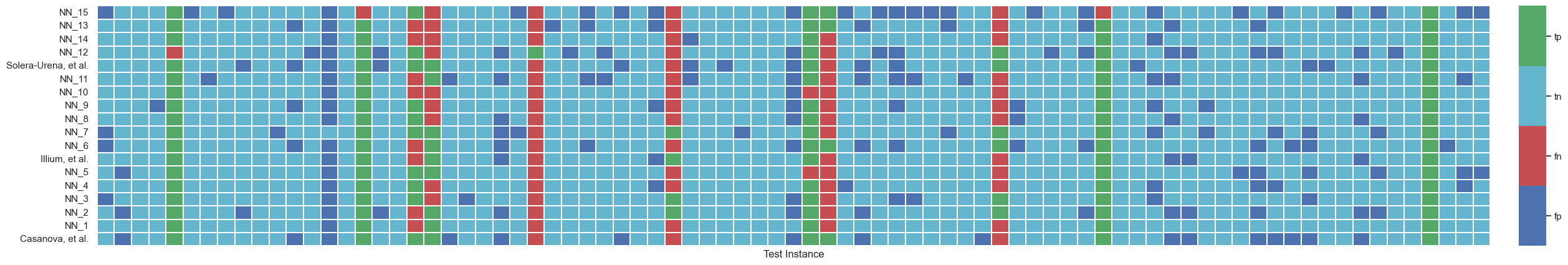}}
    \caption{Schematic detailing the level of agreement between teams for each test instance for the \textbf{COVID-19 Cough Sub-Challenge}. Each row represents a team's submission results. The teams have been ordered by Unweighted Average Recall, from the bottom up (team \textit{Casanova, et al.}'s predictions represent the highest scoring submission). Each column represents all teams predictions, across the competition, for one test instance. The test instances appear in the order in which they are in the test set. \ref{sublable1} details all the test instances, \ref{sublable2} details only the test instances which were experiencing symptoms at the time of recording, and \ref{sublable3} details only the test instances which were experiencing no symptoms at the time of recording.}

	\label{fig:agreementcough}
\end{figure}
\begin{figure}[h]
	\centering
	\subfloat[]{\label{agreementspeechsublable1}\includegraphics[width=1.4\textheight]{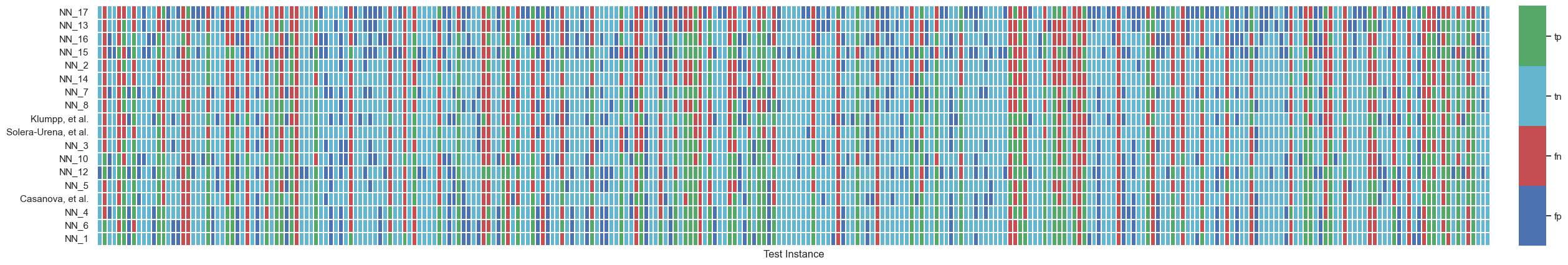}} \\
    \subfloat[]{\label{agreementspeechsublable2}\includegraphics[width=1.4\textheight]{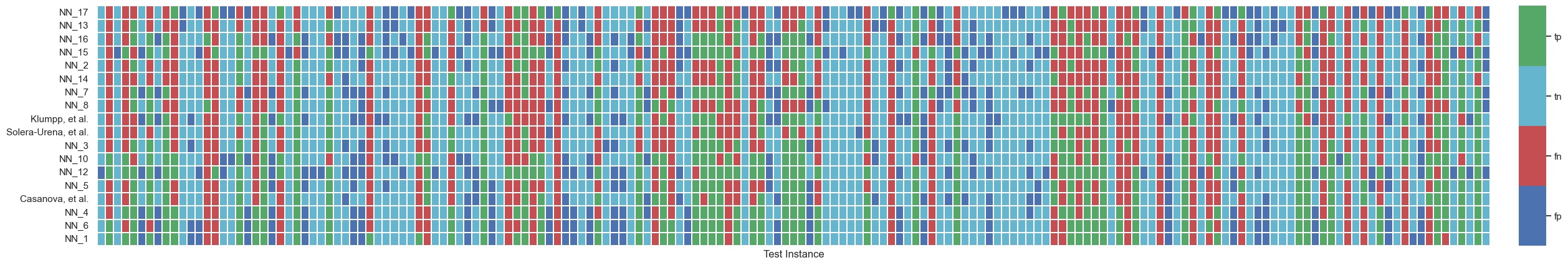}} \\
    \subfloat[]{\label{agreementspeechsublable3}\includegraphics[width=1.4\textheight]{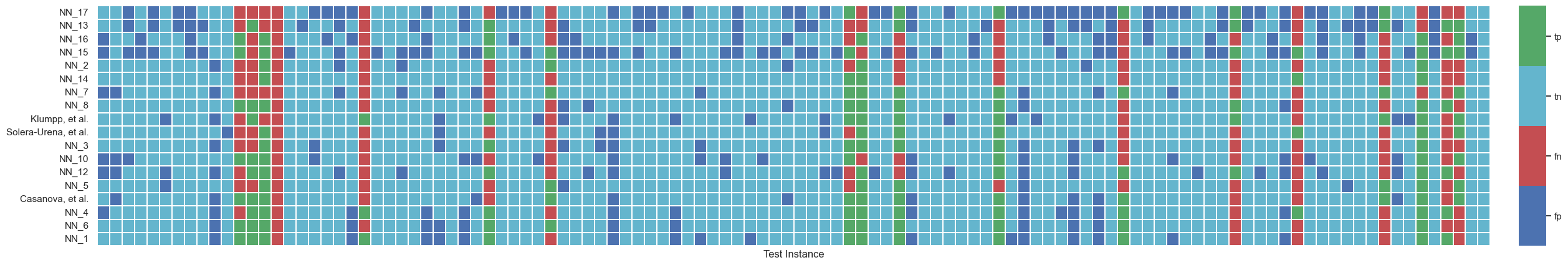}}
    \caption{Schematic detailing the level of agreement between teams for each test instance for the \textbf{COVID-19 Speech Sub-Challenge}. Each row represents a team's submission results. The teams have been ordered by Unweighted Average Recall (UAR), from the bottom up (team \textit{yoshiharuyamamoto}'s predictions represent the highest scoring submission). Each column represents all teams' predictions, across the competition, for one test instance. The test instances appear in the order which they are in the test set. note: There are more test cases in the COVID-19 Speech Sub-Challenge than in the COVID-19 Cough Sub-Challenge. \ref{agreementspeechsublable1} details all the test instances, \ref{agreementspeechsublable2} details only the test instances which were experiencing symptoms at the time of recording, and \ref{agreementspeechsublable3} details only the test instances which were experiencing no symptoms at the time of recording.}

	\label{fig:agreementspeech}
\end{figure}
\end{landscape}
\section{Acknowledgements}
We acknowledge funding from the DFG’s Reinhart Koselleck project No. 442218748 (AUDI0NOMOUS) and the ERC project No. 833296 (EAR).
\bibliographystyle{abbrvnat}

\clearpage
\section{Appendix}

\begin{figure}[h]
	\centering
	\subfloat[]{\label{controlsublable1}\includegraphics[width=0.81\textwidth]{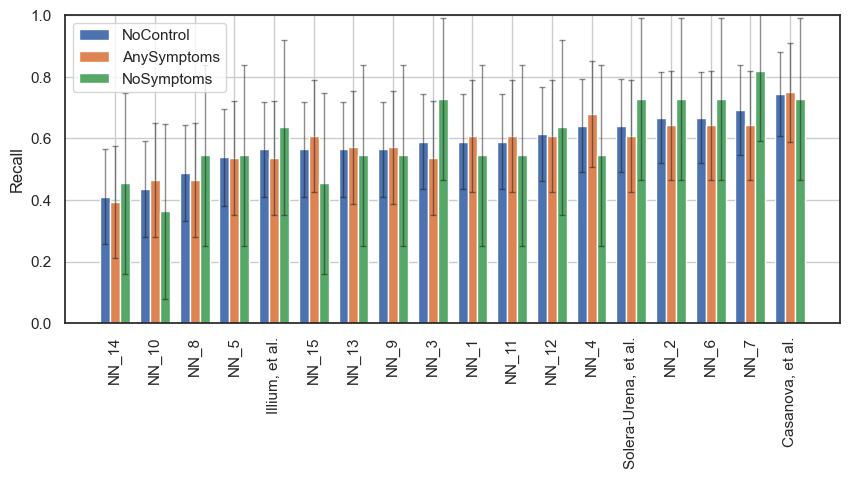}} \\
    \subfloat[]{\label{controlsublable2}\includegraphics[width=0.81\textwidth]{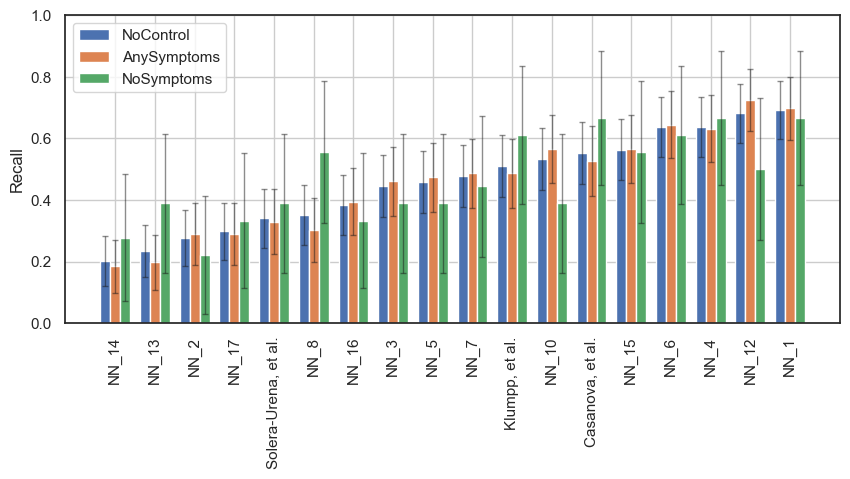}}
    \caption{Team performance on the full test set (NoControl) and two curated test sets featuring only test instances where the participants either had at least one symptom (AnySymptoms) or were displaying no symptoms at all (NoSymptoms). The metric reported is recall for positive cases. 95\,\% confidence intervals are shown, calculated via the normal approximation method. \ref{controlsublable1} corresponds to the COVID-19 Cough Sub-Challenge, CCS, and \ref{controlsublable1} the COVID-19 Speech Sub-Challenge, CSS.}
	\label{fig:control}
\end{figure}

\begin{figure}[h]
	\centering
    \includegraphics[width=0.9\textwidth]{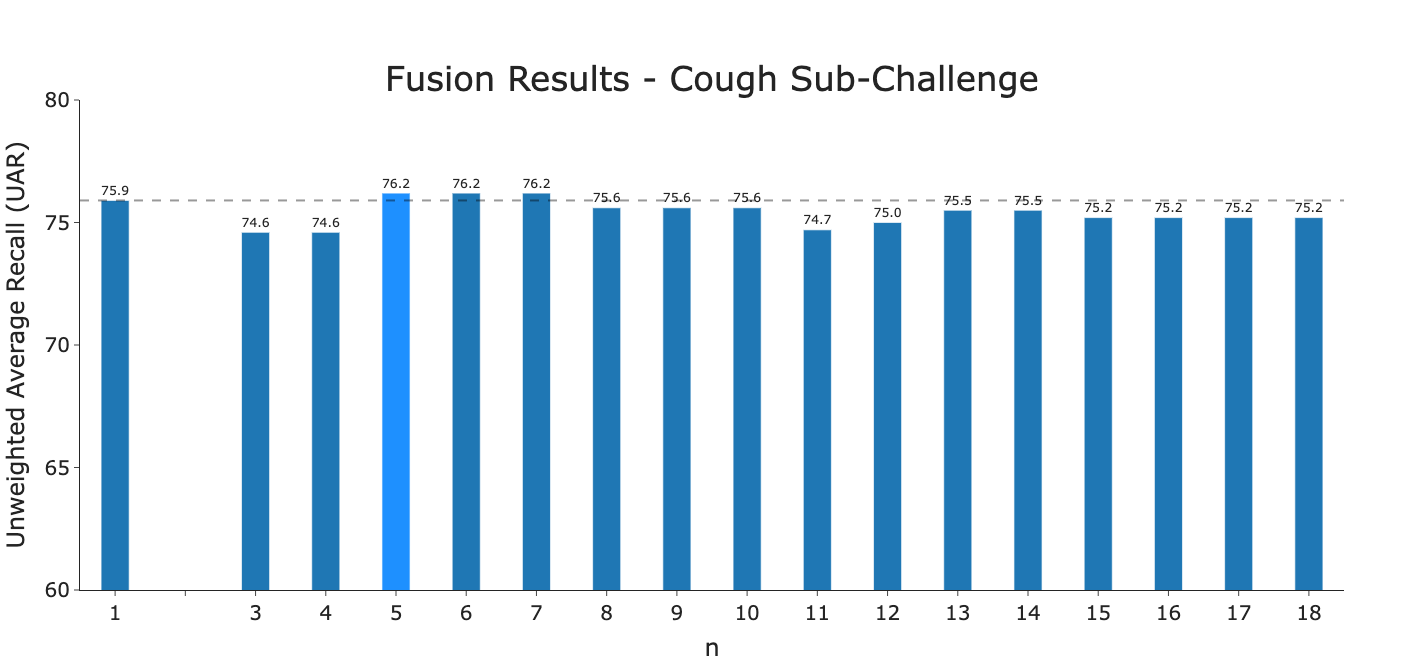}
	\caption{The performance of the fusion model of $n$-best models for the COVID-19 Cough Sub-Challenge using majority voting.}
	\label{fig:fusioncough}
\end{figure}

\begin{figure}[h]
	\centering
    \includegraphics[width=0.9\textwidth]{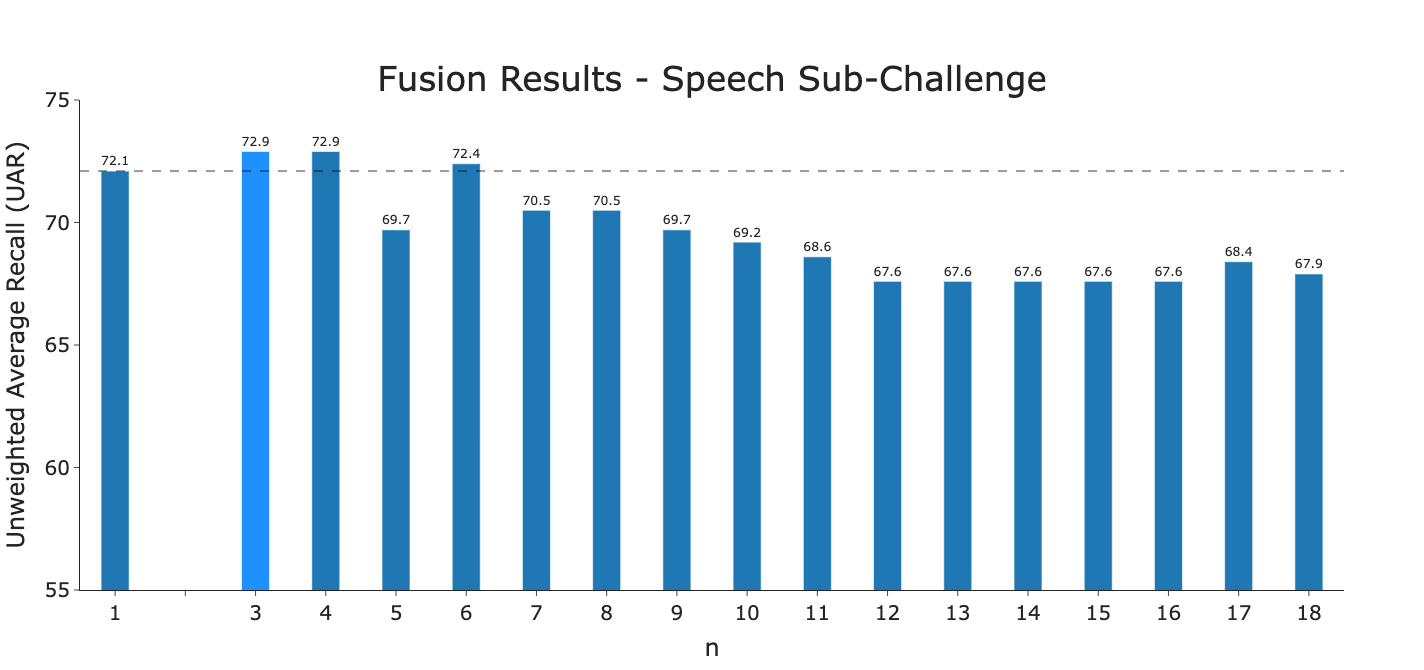}
	\caption{The performance of the fusion model of $n$-best models for the COVID-19 Speech Sub-Challenge using majority voting..}
	\label{fig:fusionspeech}
\end{figure}

\begin{landscape}
\begin{figure}
	\centering
	\subfloat[]{\label{samplerateagreementcough1}\includegraphics[width=1.4\textheight]{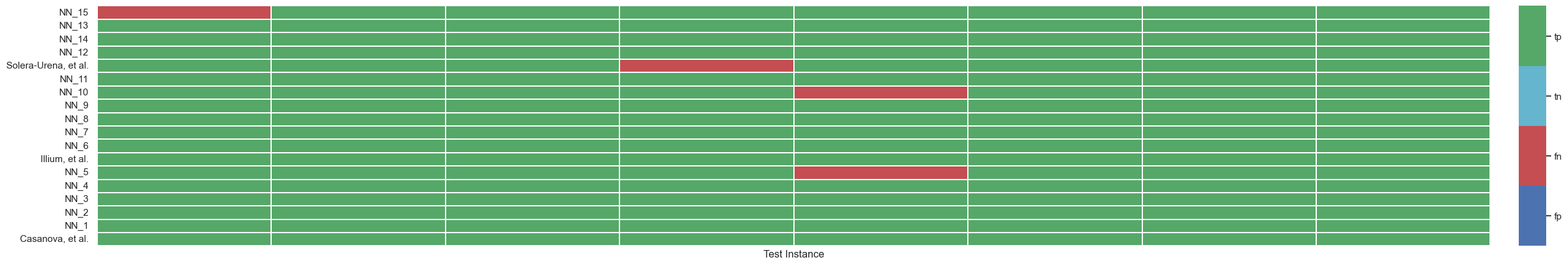}} \\
    \subfloat[]{\label{samplerateagreementcough2}\includegraphics[width=1.4\textheight]{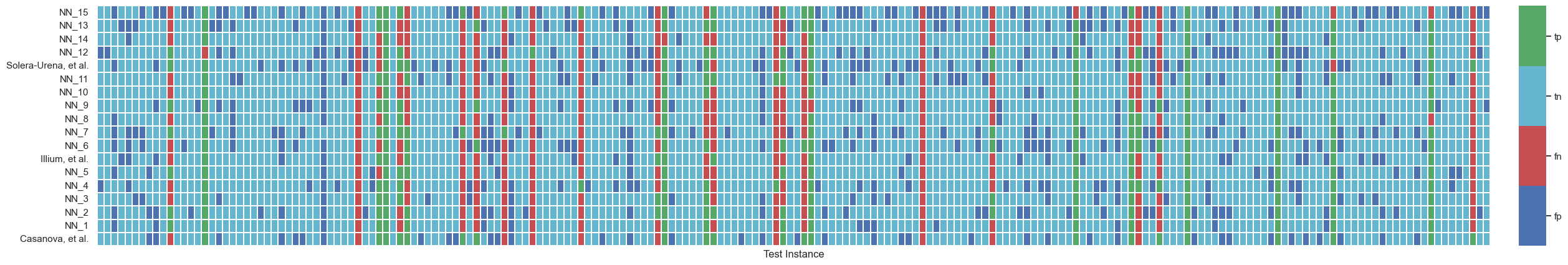}}
    \caption{Schematic detailing the level of agreement as in \ref{fig:agreementcough} with test instances with either a low sample rate (below 12\,kHz) (\ref{samplerateagreementcough1}) or high sample rate (above 12\,kHz) (\ref{samplerateagreementcough2}).}

	\label{fig:agreementcoughsampleratecough}
\end{figure}
\end{landscape}

\begin{landscape}
\begin{figure}
	\centering
	\subfloat[]{\label{samplerateagreementspeech1}\includegraphics[width=1.4\textheight]{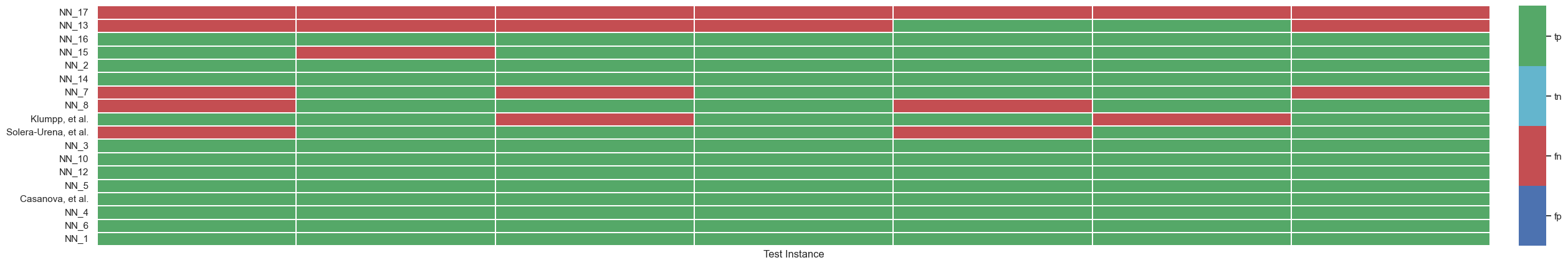}} \\
    \subfloat[]{\label{sublablerateagreementspeech2}\includegraphics[width=1.4\textheight]{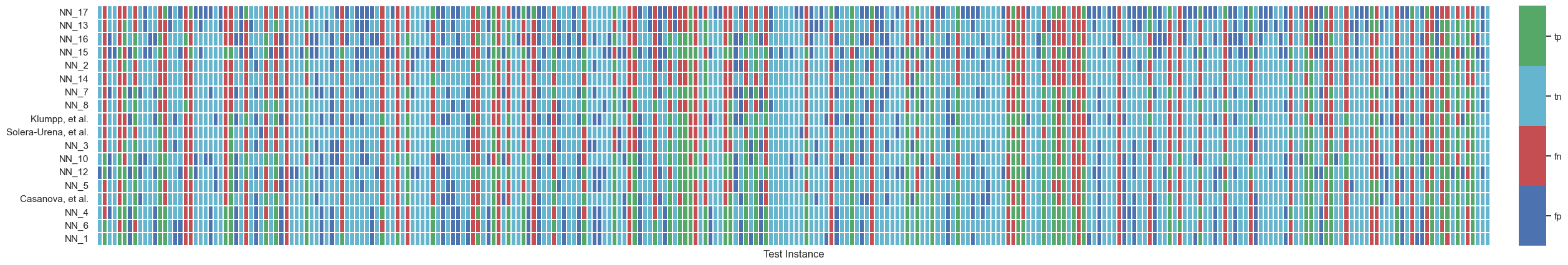}}
    \caption{Schematic detailing the level of agreement as in \ref{fig:agreementspeech} with test instances with either a low sample rate (below 12\,kHz) (\ref{samplerateagreementspeech1}) or high sample rate (above 12\,kHz) (\ref{sublablerateagreementspeech2}).}

	\label{fig:agreementspeechsampleratecough}
\end{figure}
\end{landscape}

\begin{figure}[h]
	\centering
	\subfloat[]{\label{controlsampleratesublable1cough}\includegraphics[width=0.83\textwidth]{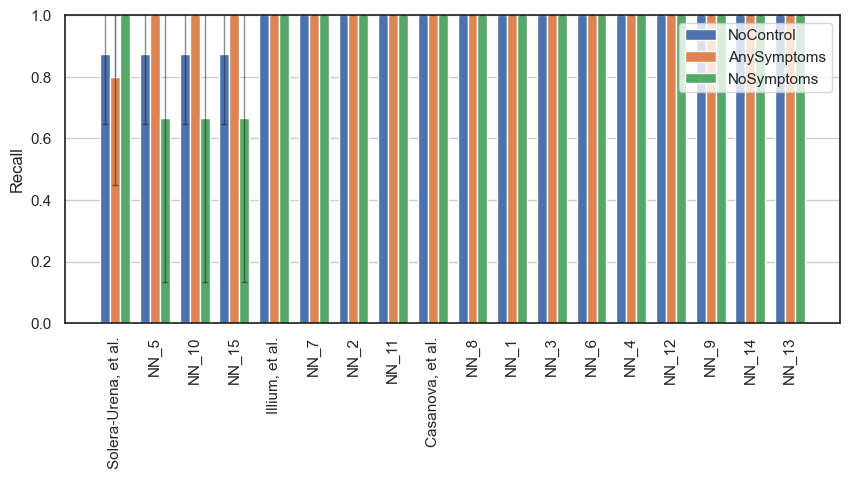}} \\
    \subfloat[]{\label{controlsampleratesublable2cough}\includegraphics[width=0.83\textwidth]{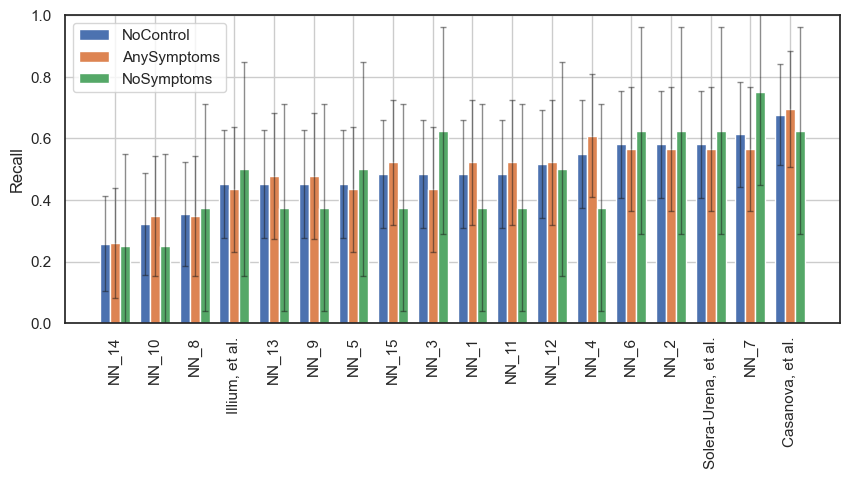}}
    \caption{Team performance on two curated test sets from the COVID-19 Cough Sub-Challenge. \ref{controlsampleratesublable1cough} controls for test samples with a sample rate of greater than 12\,kHz and \ref{controlsampleratesublable2cough} controls for test samples with a sample rate of 12\,kHz and below. The metric reported is recall for positive cases. 95\,\% confidence intervals are shown, calculated via the normal approximation method.}
	\label{fig:controlsampleratecough}
\end{figure}

\begin{figure}[h]
	\centering
	\subfloat[]{\label{controlsampleratesublable1speech}\includegraphics[width=0.83\textwidth]{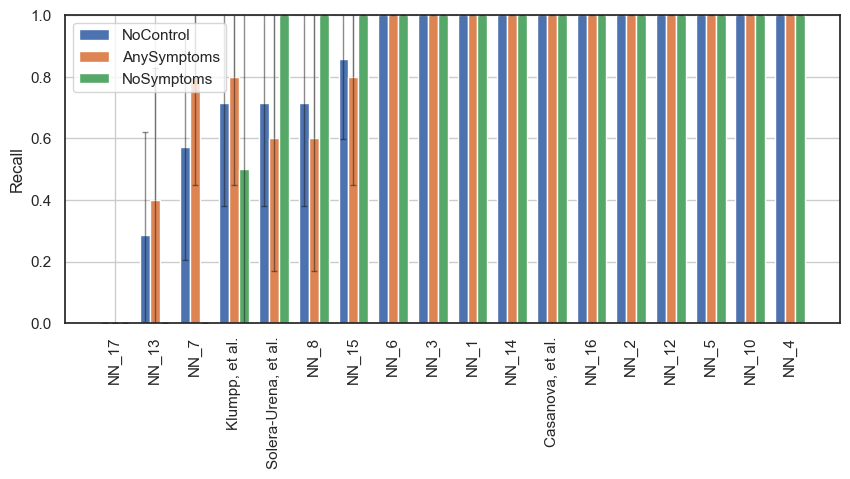}} \\
    \subfloat[]{\label{controlsampleratesublable2speech}\includegraphics[width=0.83\textwidth]{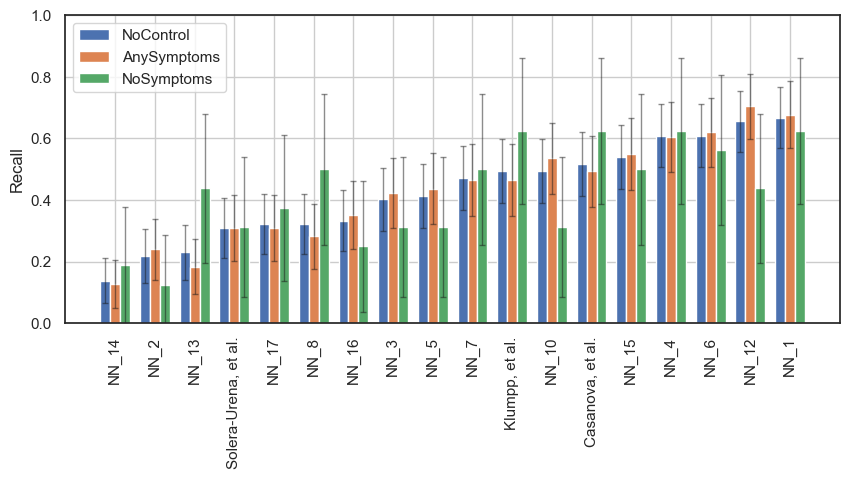}}
    \caption{Team performance on two curated test sets from the COVID-19 Speech Sub-Challenge. \ref{controlsampleratesublable1speech} controls for test samples with a sample rate of greater than 12\,kHz and \ref{controlsampleratesublable2speech} controls for test samples with a sample rate of 12\,kHz and below. The metric reported is recall for positive cases. 95\,\% confidence intervals are shown, calculated via the normal approximation method.}
	\label{fig:controlsampleratespeech}
\end{figure}

Here we present some results from ablation studies of teams' performances through evaluating performance on curated subsets of the test set. Figure \ref{fig:control} details the effect of controlling for symptom cofounders on teams' performance. Figures \ref{fig:controlsampleratecough} and \ref{fig:controlsampleratespeech} repeats this analysis however controlling for sample rate. Figures \ref{fig:agreementcoughsampleratecough} and \ref{fig:agreementspeechsampleratecough} details the level of agreement between teams for the low \ref{samplerateagreementcough1}, \ref{samplerateagreementspeech1} and high \ref{samplerateagreementcough2} \ref{sublablerateagreementspeech2} sample rate test cases. Figures \ref{fig:fusioncough} and \ref{fig:fusionspeech} detail the classification performance of a fusion of teams' predictions on the test set.

\end{document}